\documentclass[aps,pra,twocolumn,amsmath,amssymb,superscriptaddress]{revtex4}

\newcommand{\ket}[1]{|#1\rangle}

\usepackage[pdftex]{graphicx}
\usepackage{mathrsfs}

\begin{document}

\bibliographystyle{apsrev}

\title{Entanglement dynamics and quasi-periodicity in discrete quantum walks}

\author{Peter P. Rohde}
\email[]{dr.rohde@gmail.com}
\homepage{http://peterrohde.wordpress.com}
\affiliation{Centre for Quantum Computation and Communication Technology, School of Mathematics and Physics, University of Queensland, QLD 4072, Australia}

\author{Alessandro Fedrizzi}
\affiliation{Centre for Quantum Computation and Communication Technology, School of Mathematics and Physics, University of Queensland, QLD 4072, Australia}
\affiliation{Centre for Engineered Quantum Systems, School of Mathematics and Physics, University of Queensland, QLD 4072, Australia}

\author{Timothy C. Ralph}
\affiliation{Centre for Quantum Computation and Communication Technology, School of Mathematics and Physics, University of Queensland, QLD 4072, Australia}

\date{\today}

\frenchspacing

\begin{abstract}
We study the entanglement dynamics of discrete time quantum walks acting on bounded finite sized graphs. We demonstrate that, depending on system parameters, the dynamics may be monotonic, oscillatory but highly regular, or quasi-periodic. While the dynamics of the system are not chaotic since the system comprises linear evolution, the dynamics often exhibit some features similar to chaos such as high sensitivity to the system's parameters, irregularity and infinite periodicity. Our observations are of interest for entanglement generation, which is one primary use for the quantum walk formalism. Furthermore, we show that the systems we model can easily be mapped to optical beamsplitter networks, rendering experimental observation of quasi-periodic dynamics within reach.
\end{abstract}

\maketitle

\section{Introduction}

Quantum walks (QWs) \cite{bib:ADZ,bib:AAKV,bib:Childs09,bib:Lovett10,bib:Kempe08} have emerged as an interesting sub-field of quantum information processing \cite{bib:NielsenChuang00}. Most existing work into QWs has focused on their applicability to implementing quantum algorithms \cite{bib:SKW,bib:DW,bib:Childs09,bib:Lovett10}. However, several authors have also considered their ability to generate entangled states \cite{bib:VenegasAndraca09,bib:Omar06,bib:Pathak07,bib:Goyal10}, which may subsequently act as a resource for other quantum information processing protocols. In this paper we expand on previous studies into the entanglement dynamics of QWs. We show that in some circumstances regular oscillations or elegant monotonic behaviour emerges, whereas for some system parameters, highly irregular, quasi-periodic behaviour emerges in the entanglement metric of the system. We focus on bounded linear graphs with reflecting boundary conditions and show that reflecting boundaries are key in the emergence of quasi-periodic dynamics. We primarily focus on single walkers, but also provide some discussion on the case of multiple walkers and post-selection. The framework we present could easily be applied to analysing arbitrary graphs with any number of walkers. We pay special attention to the optical implementation of such QWs.

Carneiro \emph{et al.} \cite{bib:Carneiro05} previously examined in detail entanglement dynamics in discrete QWs acting on regular graphs. In this work we build on this by focusing on the quasi-periodic nature of these dynamics. While highly irregular dynamics emerge in some circumstances, the dynamics are strictly non-chaotic since the system has no non-linearities. W\'ojcik \emph{et al.} \cite{bib:Wojcik04} previously gave consideration of the quasi-periodic nature of QWs. Here we focus on entanglement dynamics with static system parameters and discuss in what circumstances regular periodic or irregular quasi-periodic behaviour emerge. We find that even with fixed, non-periodic parameters, i.e. no time-dependent or periodic choices in the coins, quasi-periodic or perfect periodic dynamics may emerge.

Considering bounded graphs is physically motivated as numerous experimental groups are beginning to experimentally demonstate QWs \cite{bib:bouwmeester_optical_1999,bib:Do05,bib:zaehringer2009rqw,schmitz2009qwt,MichalKarski07102009,bib:Schreiber10,bib:Broome10,bib:Peruzzo10}, but clearly infinite graphs are not possible, so considering the effects of boundary conditions is relevant to such studies. Furthermore, we discuss the mapping of our QW formalism to the specific case of photonic implementation, which is relevant to present optical experiments \cite{bib:Schreiber10,bib:Broome10,bib:Peruzzo10}.

\section{Background}

A QW comprises a system of \emph{positions} (vertices in a graph), each of which may be occupied by a \emph{walker} (e.g. a particle). A QW is defined by couplings between position states on some graph $G$. A position state may literally refer to a physical position, such as a spatial mode in optics, or more abstractly to a level in a higher dimensional system. The walker is an excitation, such as a photon, which may occupy a position, and is a bipartite system consisting of a position, $x$, and a \emph{coin} value, $c$, $\ket{x,c}$. The coin is a parameter that indicates the direction the walker will take through the graph at the next step. The archetypal QW consists of two operations, the \emph{coin} operator and the \emph{step} operator. The coin operator coherently specifies the coin value, which in general may be a superposition, while the step operator updates the position of the walker based on the coin state. This is in contrast to a classical random walk where the coin operator probabilistically specifies the coin value.

The standard illustrative example of a discrete-time QW is that of a single walker evolving along an infinitely long linear graph structure. In this case we can define the coin and step operators as,
\begin{eqnarray} \label{eq:linear_graph_one_walker}
C\ket{x,\pm 1} &=& (\ket{x,-1} \pm \ket{x,1})/\sqrt{2} \nonumber \\
S\ket{x,c} &=& \ket{x+c,c},
\end{eqnarray}
where $C$ is the Hadamard coin in this example. After $t$ time steps the evolution of the system is given by $(SC)^t$. In general the coin operator may vary at each time step in which case the evolution is of the form \mbox{$\prod_{i=1}^t SC_i$}.

\subsection{Walker operator formalism}

The simple illustrative example presented above is appropriate when a single walker is present and the graph has infinite linear structure (see for example Ref. \cite{bib:Lovett10}). However that formalism becomes unwieldy for multiple walkers acting on general graphs. Thus we turn to the \emph{walker operator} formalism presented by Rohde \emph{et al.} \cite{bib:RohdeSchreiber10}. Here a walker creation operator, $w(x,c)^\dag$, acts on an empty system $\ket{0}$, to create a walker with given position and coin values,
\begin{equation}
w(x,c)^\dag\ket{0} = \ket{x,c}.
\end{equation}
In the photonic case $\ket{0}$ corresponds to the vacuum state. The state of a single walker system on graph $G$ can be written in the form
\begin{equation}
\sum_{i,j \in G} \alpha_{ij} w(i,j)^\dag \ket{0}.
\end{equation}
Evidently, with the interpretation that a walker is a particle occupying distinct physical positions \footnote{As opposed to the interpretation that position states are positions in Hilbert space.}, the state is always of a biased W-type \cite{bib:Duer00}, i.e. a biased superposition of a single excitation across a number of positions. Thus, the class of entangled states that may be generated is rather restrictive \footnote{Historically, there has been some debate as to whether a single walker (e.g. photon) state can in fact be entangled. We take the current view that it can be, since if we couple each mode in the system to an atom (i.e. local operations) we can create a superposition of a single excitation across multiple atoms, which is entangled.}, and does not include some states that are of practical interest such as GHZ \cite{bib:GHZ89} and cluster (or graph) states \cite{bib:Raussendorf01,bib:Raussendorf03}. However, W-states are nonetheless of practical interest since they are more resilient against loss than other entanglement classes, and have applications in quantum cryptography \cite{bib:BB84}, teleportation \cite{bib:Bennett93,bib:Shi02,bib:Joo03} and computation \cite{bib:NielsenChuang00}.

\begin{figure}[!htb]
\includegraphics[width=0.6\columnwidth]{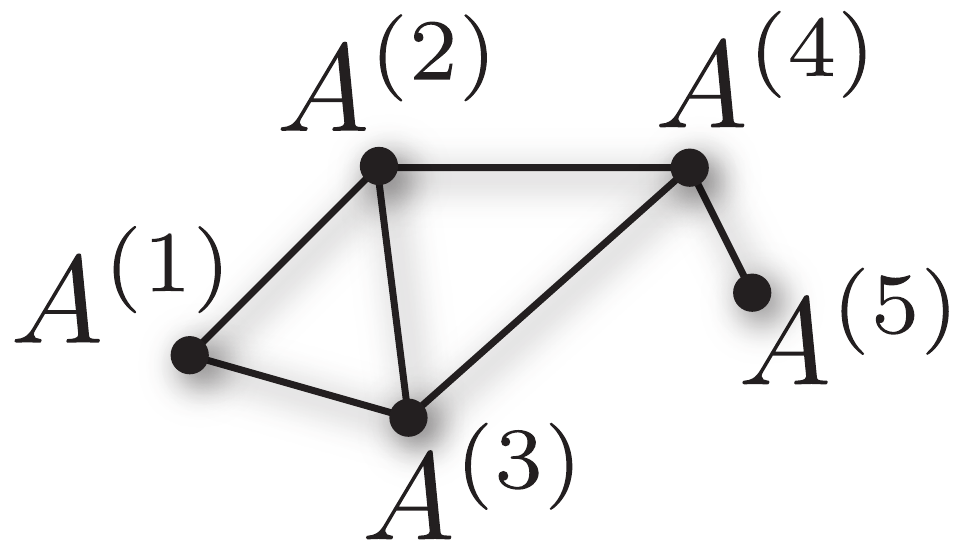}
\caption{An example walker graph with five position states. At each position $i$, a local operation $A^{(x)}$ is applied to $x$ and its neighbours.} \label{fig:example_graph}
\end{figure}

In analogy with the conventional description of a QW, we define the evolution of the system using \emph{coin} and \emph{step} operators. The coin operator flips a quantum mechanical coin which determines the direction of the walker. The step operator then uses the coin value to propagate the walker.
\begin{eqnarray} \label{eq:general_coin_step_pd}
&C:& \, w(x,c)^\dag \mapsto \sum_{j\in n_x} A_{cj}^{(x)} w(x,j)^\dag \nonumber \\
&S:& \, w(x,j)^\dag \mapsto w(j,x)^\dag.
\end{eqnarray}
Here $n_x$ is the neighbourhood of position $x$ in $G$, which in general may be a directed graph \footnote{The neighbourhood of $x$ is defined as the set of vertices connected to $x$ via an edge (an outgoing edge in the case of a directed graph).}, and $A^{(x)}$ are local operators acting on the respective neighbourhood. Thus, the walker may only ever evolve to positions within its neighbourhood in the graph. An example of a walker graph is shown in Fig. \ref{fig:example_graph}. Importantly, the above formalism allows for the modelling of irregular graphs, i.e. graphs where the degree varies at different vertices. Note the subtle difference in the definition of the step operator between the formalisms presented in Eqs. \ref{eq:linear_graph_one_walker} and \ref{eq:general_coin_step_pd}. In the former the step operator does not alter the coin state, whereas in the latter both the position and coin state are updated. This slight change in formalism is adopted to allow for arbitrary graphs to be described.

\begin{figure}[!htb]
\includegraphics[width=0.7\columnwidth]{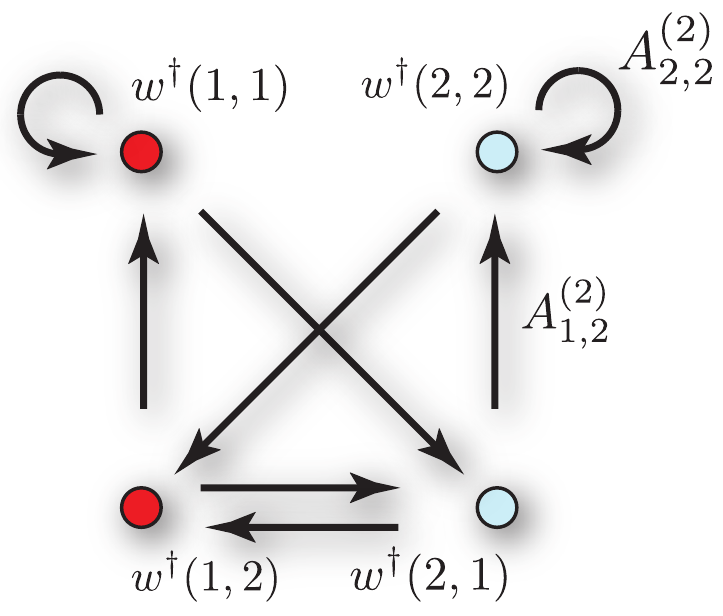}
\caption{(Colour online) The allowed transitions between walker operators for two position states. Vertices in red are associated with position $1$, while vertices in blue are associated with position $2$.} \label{fig:transitions}
\end{figure}

Once the coin operator is applied, the coin parameter can be interpreted as the next position of the walker, while after the step operation it can be regarded as a memory of the walker's previous position value. These properties are necessary to ensure unitarity of the operations. In practise it is often somewhat artificial to consider the coin and step as separate operators; they can be merged into a combined evolution operator,
\begin{equation}
E=SC: \, w(x,c)^\dag \mapsto \sum_{j\in n_x} A_{cj}^{(x)} w(j,x)^\dag.
\end{equation}
From this formalism one can express the allowed walker operator transitions as a graph. An example of the transition graph for a two-position system is shown in Fig. \ref{fig:transitions}.

\subsection{Global system evolution}

Based on the local neighbourhood operations we can construct the unitary operator acting on the entire system. For a single walker QW we can write a system of coupled, discrete time-series equations as
\begin{equation}
\alpha_{ij}(t+1) = \sum_k \alpha_{jk}(t) A_{ki}^{(j)}.
\end{equation}
This can be rewritten in matrix form as
\begin{equation} \label{eq:linear_map}
\vec{\alpha}(t+1) = U \vec{\alpha}(t),
\end{equation}
where
\begin{equation}
U_{ij,kl} = A_{li}^{(j)} \delta_{kj}.
\end{equation}
As an example, for a single walker, two-position system with balanced Hadamard coins,
\begin{equation}
C_H = \frac{1}{\sqrt{2}} \left( \begin{array}{cc}
1 & 1 \\
1 & -1 \\
\end{array} \right),
\end{equation}
the global evolution is given by
\begin{equation}
U =  \frac{1}{\sqrt{2}} \left(
\begin{array}{cccc}
 1 & 1 & 0 & 0 \\
 0 & 0 & 1 & 1 \\
 1 & -1 & 0 & 0 \\
 0 & 0 & 1 & -1
 \end{array}
\right),
\end{equation}
and for a single walker, three-position system by,
\begin{equation}
U = \frac{1}{\sqrt{2}} \left(
\begin{array}{ccccccccc}
 1 & 1 & 0 & 0 & 0 & 0 & 0 & 0 & 0 \\
 0 & 0 & 0 & 1 & 0 & 1 & 0 & 0 & 0 \\
 0 & 0 & 0 & 0 & 0 & 0 & \sqrt{2} & 0 & 0 \\
 1 & -1 & 0 & 0 & 0 & 0 & 0 & 0 & 0 \\
 0 & 0 & 0 & 0 & \sqrt{2} & 0 & 0 & 0 & 0 \\
 0 & 0 & 0 & 0 & 0 & 0 & 0 & 1 & 1 \\
 0 & 0 & \sqrt{2} & 0 & 0 & 0 & 0 & 0 & 0 \\
 0 & 0 & 0 & 1 & 0 & -1 & 0 & 0 & 0 \\
 0 & 0 & 0 & 0 & 0 & 0 & 0 & 1 & -1
\end{array}
\right).
\end{equation}

\subsection{Optical implementation}

\begin{figure}[!htb]
\includegraphics[width=0.7\columnwidth]{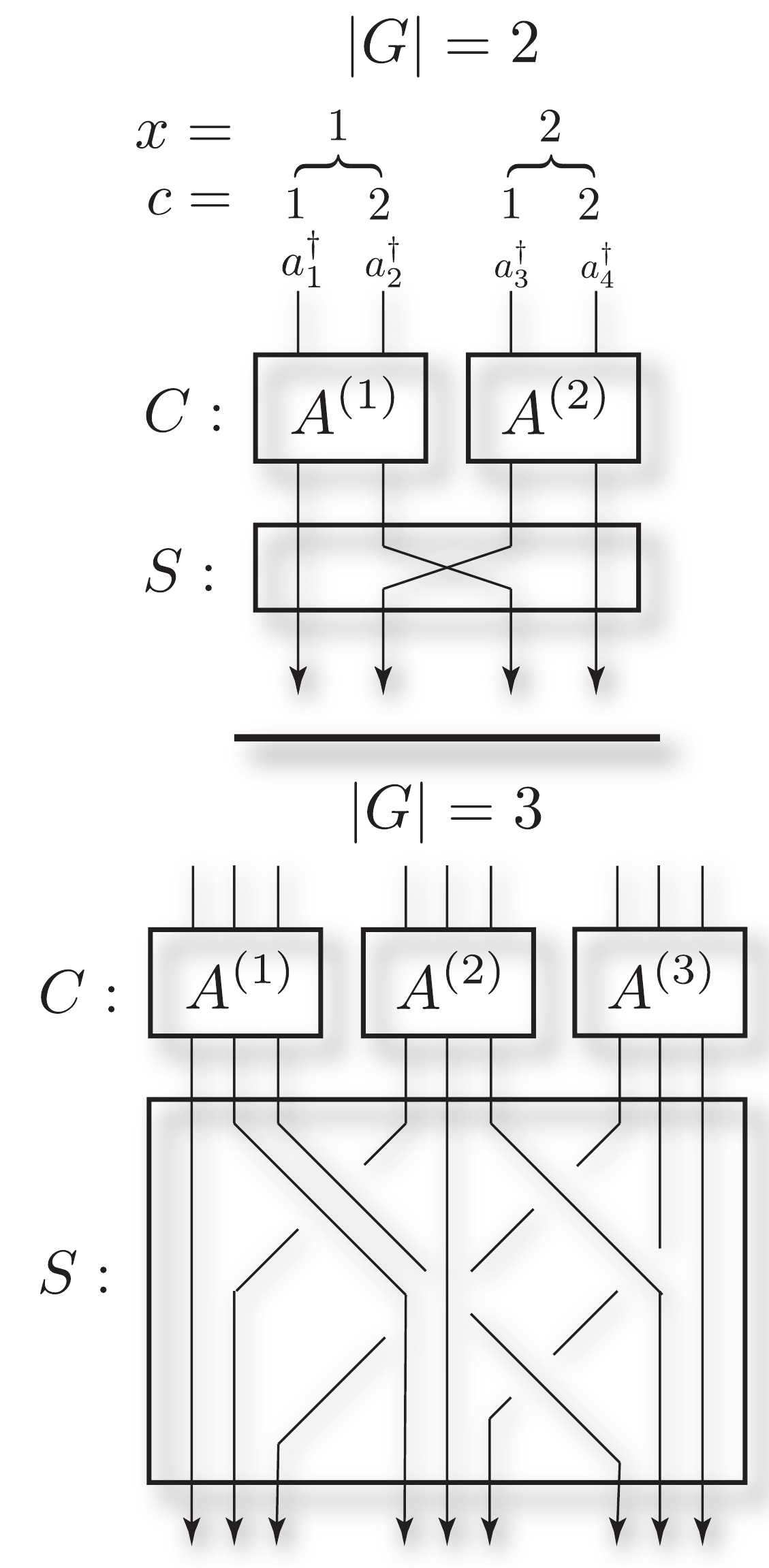}
\caption{Mapping a two- (top) and a three-position walk (bottom) to a coinless photonic implementation. The coin operator ($C$) applies the unitary evolution $A^{(i)}$ to each `bundle' of modes corresponding to a given position $i$. The step operator ($S$) then simply permutes spatial modes appropriately.} \label{fig:mapping}
\end{figure}

We now turn our attention to the specific case of optical implementation using single photons. In the optical case the described formalism is a little detached from the physical architecture as photons may have a well defined position, via distinct spatial modes, but have no intrinsic coin (i.e. memory) state \footnote{In present day experiments with a walker on a line the photon may have a coin state via the polarisation degree of freedom. However, polarisation is a two-level Hilbert space. Thus, for higher order graphs where the number of available coin states is $|G|$, polarisation is insufficient for use as a coin state. Orbital angular momentum modes are a candidate for use as coin states, however this is experimentally challenging.}. To overcome this limitation we can map the walker operator formalism to a photon creation operator formalism by defining a function which maps the tuple \mbox{$\{x,c\}$} to a single number $n$ which corresponds to the photon's spatial mode, denoted by the photon creation operator $a_n^\dag$, where
\begin{equation}
n = (x-1)d + c.
\end{equation}
The reverse mapping can easily be performed using modulo arithmetic. Having performed this mapping, it is straightforward to physically implement the walker operator formalism in an optical context. Examples for mapping two- and three-position walks are shown in Fig. \ref{fig:mapping}.

\section{Entanglement dynamics}

Next we consider the entanglement dynamics of a QW. For a single walker we characterise the entanglement of the system using the Shannon entropy of the probability distribution of the walker across the position/coin basis states,
\begin{equation}
E = -\sum_{i,j\in G} |\alpha_{ij}|^2 \, \mathrm{log}_2 \, |\alpha_{ij}|^2.
\end{equation}
This metric varies between $0$, for a completely localised (i.e. unentangled) state, and \mbox{$\mathrm{log}_2 \, |G|^2$} for a maximally entangled state (i.e. balanced W-state).

For multiple walkers we can introduce the Meyer-Wallach entanglement measure \cite{bib:MeyerWallach02}, given by
\begin{equation}
E = 1- \frac{1}{|G|^2} \sum_{i=1}^{|G|^2} \mathrm{tr} (\rho_i^2),
\end{equation}
where $\rho_i$ is the reduced state of subsystem $i$. \mbox{$E=0$} for a completely separable system, and \mbox{$E>0$} is a signature of entanglement. This measure effectively averages over the purity of the reduced state of each site in the system. This was previously applied to QWs in Ref. \cite{bib:Goyal10}.

\subsection{Linear walks with a single walker}

We focus on the case of a single walker evolving through a bounded linear graph. We will study biased Hadamard coins,
\begin{equation}
H_\delta = \left( \begin{array}{cc}
\sqrt{\delta} & \sqrt{1-\delta} \\
\sqrt{1-\delta} & -\sqrt{\delta} \\
\end{array} \right),
\end{equation}
and examine the dependence of the entanglement dynamics on the coins'  bias for varying graph sizes. For \mbox{$\delta=1$} this operator reduces to the Pauli $Z$ operator, and for \mbox{$\delta=0$} to the Pauli $X$ operator, neither of which create superpositions. However for \mbox{$0<\delta<1$} this coin creates superpositions of coin states, which subsequently creates entanglement.

\subsubsection{Balanced Hadamard coins}

\begin{figure*}[!htb]
\includegraphics[width=0.82\textwidth]{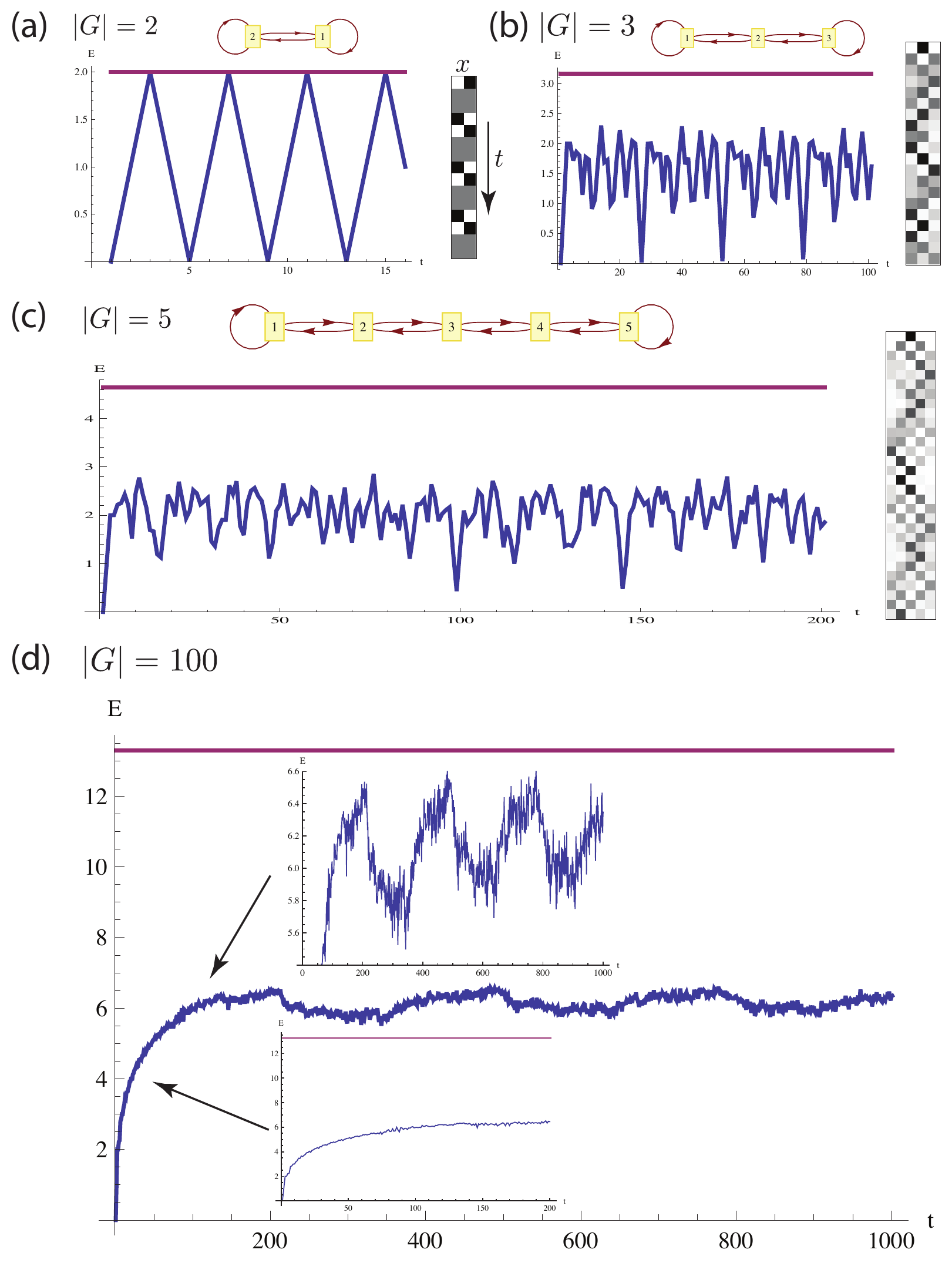}
\caption{(Colour online) Entanglement and position amplitude evolution for linear graphs of various degree, $|G|$, and a fixed balanced Hadamard coin, \mbox{$\delta=1/2$}. In the position amplitude plots, black corresponds to unit amplitude while white corresponds to zero amplitude. Blue lines show the entanglement dynamics, while red lines show the maximum possible entanglement (i.e. that of a balanced W-state). In (d) the insets show the monotonic and irregular/quasi-periodic sections in the evolution. Note the different time scales on the graphs.} \label{fig:balanced_plots}
\end{figure*}

First consider the case of balanced Hadamard coins (i.e. \mbox{$\delta=1/2$}) for linear graphs of varying lengths. Several examples of the entanglement generated on graphs of varying degrees are illustrated in Fig. \ref{fig:balanced_plots}. In the case of \mbox{$|G|=2$}, there are four basis states. The walker is initialised in a localised state and \mbox{$E=0$}. After one time step the state evolves into a Bell state across two of the basis states, after two time steps to a W-state across all four basis states (i.e. the maximum possible entanglement), after three time steps back to a Bell state, and then finally back to a localised state. Thus, the entanglement exhibits perfect periodicity with \mbox{$T=4$}. For \mbox{$|G|=3$} the entanglement metric becomes quasi-periodic, with apparent irregularity within the periods, and for \mbox{$|G|=5$} the behaviour seems to be completely irregular. For a very long graph, \mbox{$|G|=100$}, the entanglement initially increases monotonically to an asymptote since the walkers are spreading out and able to occupy an increasing number of positions. It is not until well after the walker hits the boundary of the graph that irregular behaviour emerges. Again, after this point the behaviour is quasi-periodic with highly irregular oscillations within the period. We note that in the case of \mbox{$|G|=100$}, the entanglement dynamics prior to reaching the graph's boundary mimic those observed in Ref. \cite{bib:VenegasAndraca09}, where an infinite linear graph was considered. Clearly this is expected since prior to the walker hitting the boundary, the length of the graph is irrelevant to the dynamics of the system.

\subsubsection{Biased Hadamard coins}

\begin{figure*}[!htb]
\includegraphics[width=0.8\textwidth]{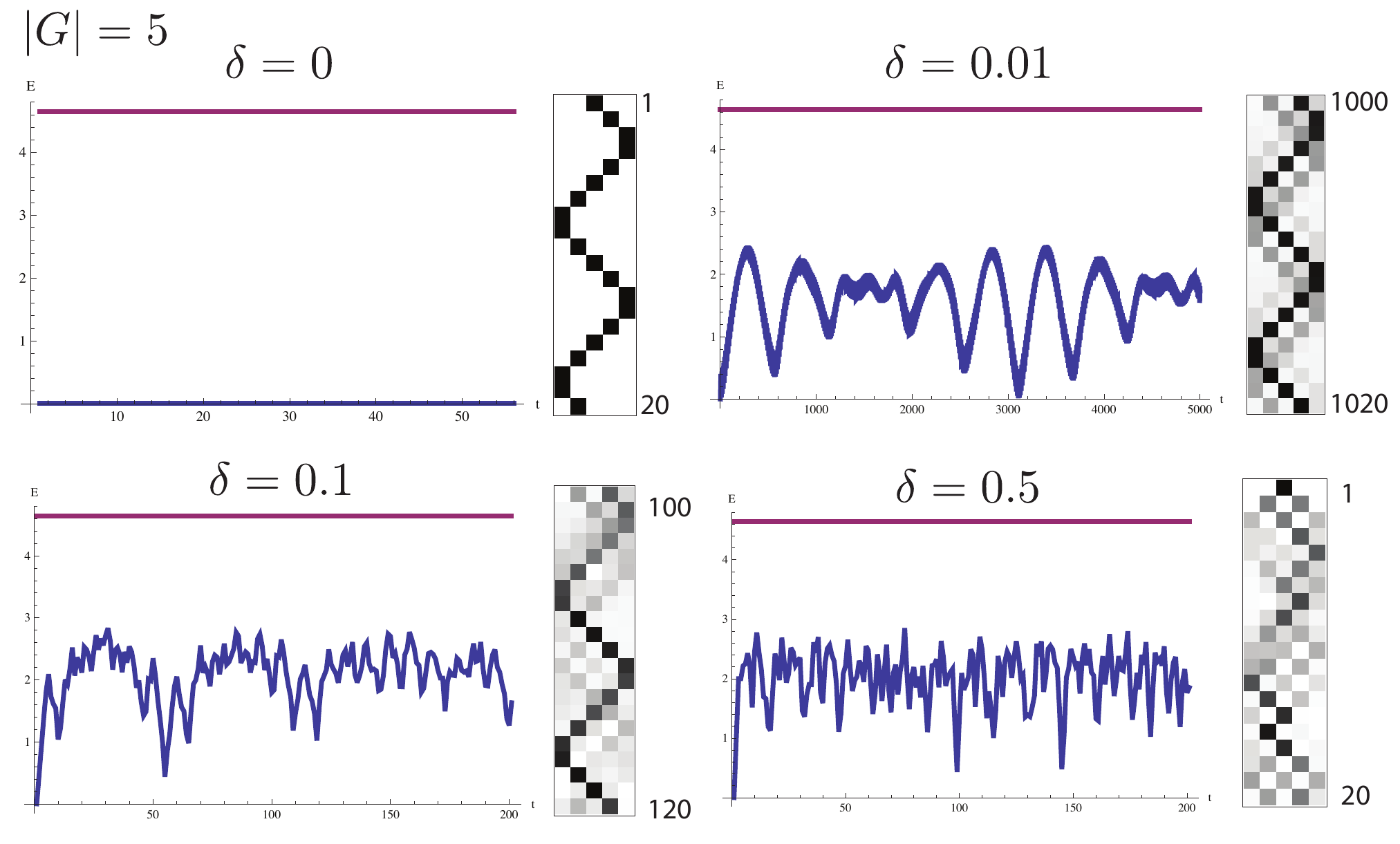}
\caption{(Colour online) Entanglement and position amplitude evolution for different coin biases, $\delta$, and fixed graph length \mbox{$|G|=5$}. Note the different time scales on the graphs.} \label{fig:biased_plots}
\end{figure*}

Next we turn our attention to the effects of coin bias in the evolution of the walk. Fig. \ref{fig:biased_plots} shows the entanglement and amplitude dynamics for a \mbox{$|G|=5$} linear graph, but with different coin biases. For \mbox{$\delta=0$} the coin does not create a superposition of coin values, thus the walker wanders around while remaining completely localised at all times, and \mbox{$E=0$}. For \mbox{$\delta=0.01$} the behaviour becomes quasi-periodic but regular, and as $\delta$ is increased further the dynamics become increasingly irregular over the illustrated time scale.

\section{Properties of the entanglement dynamics}

We now examine some of the characteristics of the entanglement dynamics of the system. Some of the features are similar to chaotic systems, however our system is not truly chaotic since the evolution of the system is linear. We consider two aspects of the system's evolution: (1) sensitivity to the system's parameters, and (2) the periodicity/quasi-periodicity of the dynamics.

\subsection{Sensitivity to system parameters}

\begin{figure}[!htb]
\includegraphics[width=\columnwidth]{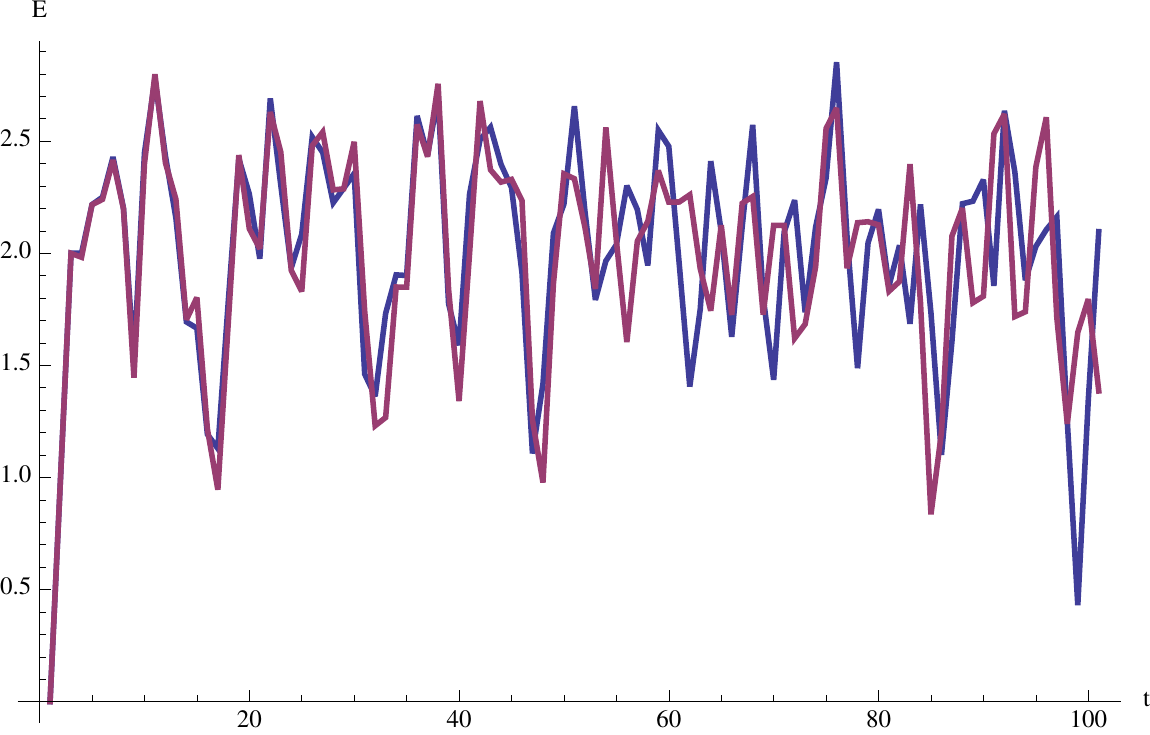}
\caption{(Colour online) Comparison of the entanglement dynamics for a \mbox{$|G|=5$} linear walk with \mbox{$\delta=0.5$} and \mbox{$\delta=0.51$}, showing the system's sensitivity to slight perturbations in the coin bias.} \label{fig:sensitivity}
\end{figure}

There are several parameters characterising our QW system: (1) the graph structure, (2) the initial condition, and (3) the coin bias. In this case we will focus on a linear graph and consider the effects of coin bias. In Fig. \ref{fig:sensitivity} we consider a \mbox{$|G|=5$} linear walk with two slightly different coin biases, \mbox{$\delta=0.5$} and \mbox{$\delta=0.51$}. Both walks have the same initial condition where the walker is initially completely localised at the middle of the graph. Evidently the dynamics exhibit high sensitivity to the different coin bias parameters. Both evolutions initially exhibit similar irregular dynamics but quickly diverge and follow different trajectories. We conclude that the entanglement dynamics in QWs are in general highly sensitive to the system's parameters.

\subsection{Periodicity}

Next we consider the periodicity of the system's evolution for different graph sizes. Revivals in QWs have previously been studied by {\v S}tefa{\v n}{\' a}k \emph{et al.} \cite{bib:Stefanak10}. Visually (see Fig. \ref{fig:biased_plots}), it is evident that the dynamics of the system are quasi-periodic, with repetitive structure over varying time scales. To analyse this more quantitatively we can consider the eigenvalue spectra of the unitary matrices describing the evolution of the systems. Because we are dealing with a closed system (i.e. the magnitude of the wave-function is always preserved --- it is never dissipative), the eigenvalues are all of unit magnitude, but in general complex. Furthermore, in general some of the eigenvalues will be degenerate. The arguments of the different eigenvalues then quantify the periodicity of the different eigenvectors of the unitary matrix. Thus the eigenvalue spectrum quantifies the periodicity of the system's evolution over different time scales and as a function of the initial state.

\begin{figure}[!htb]
\includegraphics[width=\columnwidth]{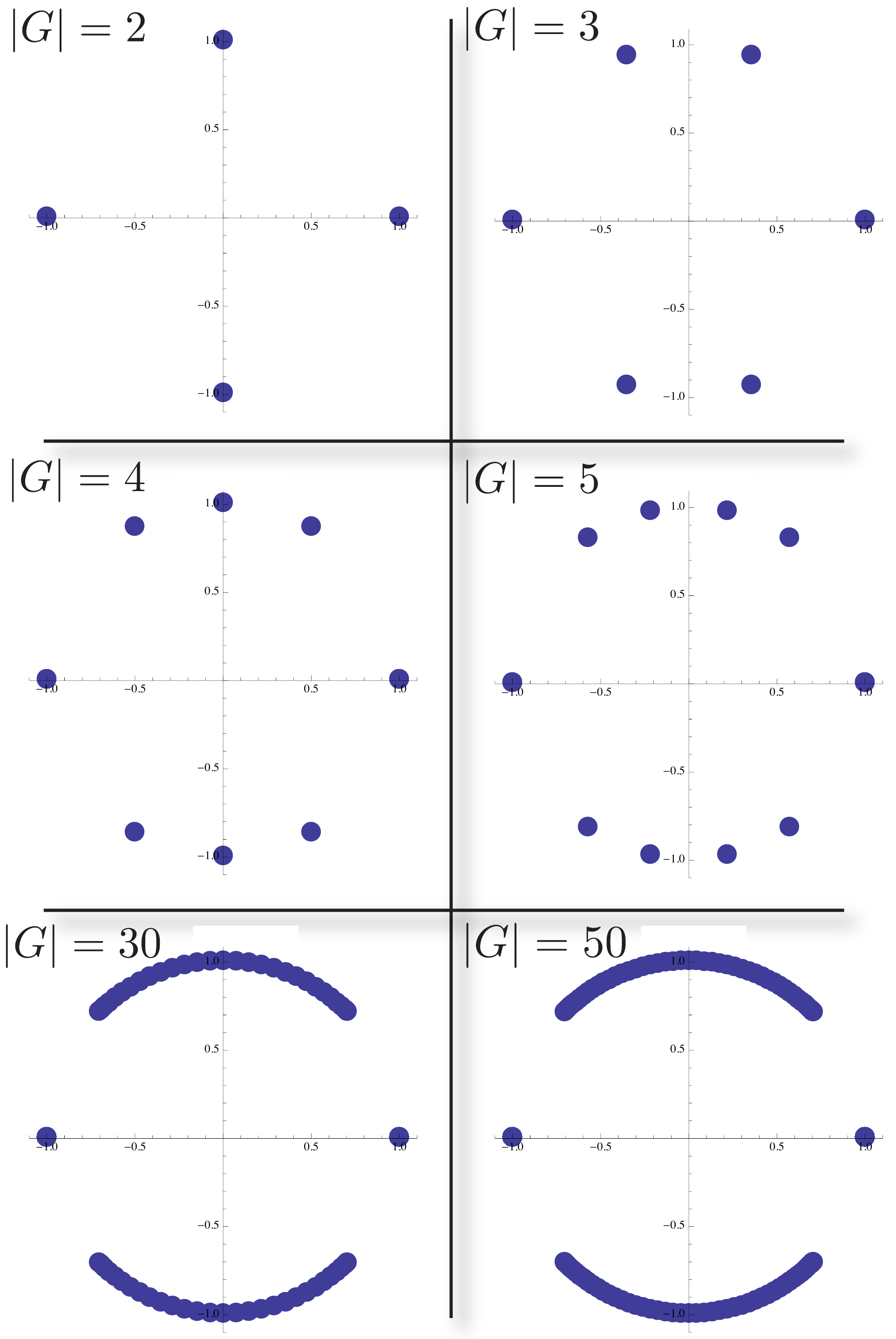}
\caption{(Colour online) Eigenvalue spectra for linear graphs of different lengths $|G|$. Note the bunching of the eigenvalues as $|G|$ is increased.} \label{fig:eigenvalues}
\end{figure}

The eigenvalue spectra for different linear graph sizes are illustrated in the complex plane in Fig. \ref{fig:eigenvalues}. Note the bunching of the eigenvalues as $|G|$ is increased. For \mbox{$|G|=2$} and \mbox{$|G|=4$} all eigenvalues have arguments of rational multiples of $\pi$, whereas for \mbox{$|G|=3$} and \mbox{$|G|=5$} at least some of the eigenvalues' arguments are irrational multiples of $\pi$. Thus, the \mbox{$|G|=2$} and \mbox{$|G|=4$} linear graphs will exhibit perfect periodicity, whereas the \mbox{$|G|=3$} and \mbox{$|G|=5$} graphs will have infinite periodicity. More generally, when the arguments of all the eigenvalues are rational, perfect periodicity will be observed, whereas when some of the arguments are irrational the system can be made to revive to within arbitrary accuracy over a sufficient timescale, although perfect periodicity will not be observed. Specifically, if the arguments of the eigenvalues can be expressed in rational form, \mbox{$\lambda_j = e^{2\pi i p_j/q_j}$}, then the periodicity of the system for a completely general state is
\begin{equation}
T = \mathrm{lcm}(t_1,\dots,t_{{|G|}^2}),
\end{equation}
where $t_j = q_j/\mathrm{gcd}(p_j,q_j)$ \footnote{If $p_j/q_j$ is already in lowest denominator form then we simply have $t_j=q_j$.}. If we relax the conditions such that the system must revive up to a global phase then the periodicity may be a fraction of $T$. Obviously in special cases, such as where the walker is initialised into a linear combination of some subset of the eigenstates of $U$, the period may be less than $T$.

\section{Multiple walkers and post-selection}

Having studied the situation of single walkers, which we have shown lead to quasi-periodic behaviour, we now turn our attention to the case of multiple walkers. Multi-walker systems are of interest, particularly in the optical context, since this is where exponential complexity in the system emerges and the possibility of exponential speedup of algorithms presents itself \cite{bib:RohdeComment10}. The power of the presented walker operator formalism is its ability to easily model higher numbers of walkers in a QW system. We show that with multiple walkers, perhaps not surprisingly, chaotic behaviour is still not possible, even with the addition of post-selection. It is well known that post-selection can induce non-linear effects in optical systems \footnote{That is, the Hamiltonian describing the evolution of the system has non-linear terms.} \cite{bib:Scheel03}, which is the basis of linear optics quantum computing (LOQC) \cite{bib:KLM01}, thus opening the possibility of chaotic behaviour, since non-linearities are a necessary condition for the emergence of chaos. We explicitly work through the case of two walkers, but the expressions can easily be generalised to more walkers.

For simplicity we will analyse the specific situation of photonic systems, since then we can ignore internal degrees of freedom (i.e. coin states) and focus just on the position state, thereby simplifying the analysis. However, as discussed previously, a walker operator formalism can always be mapped to a photon creation operator formalism. Thus there is no loss of generality in restricting ourselves to the photonic case.

\begin{figure}[!htb]
\includegraphics[width=\linewidth]{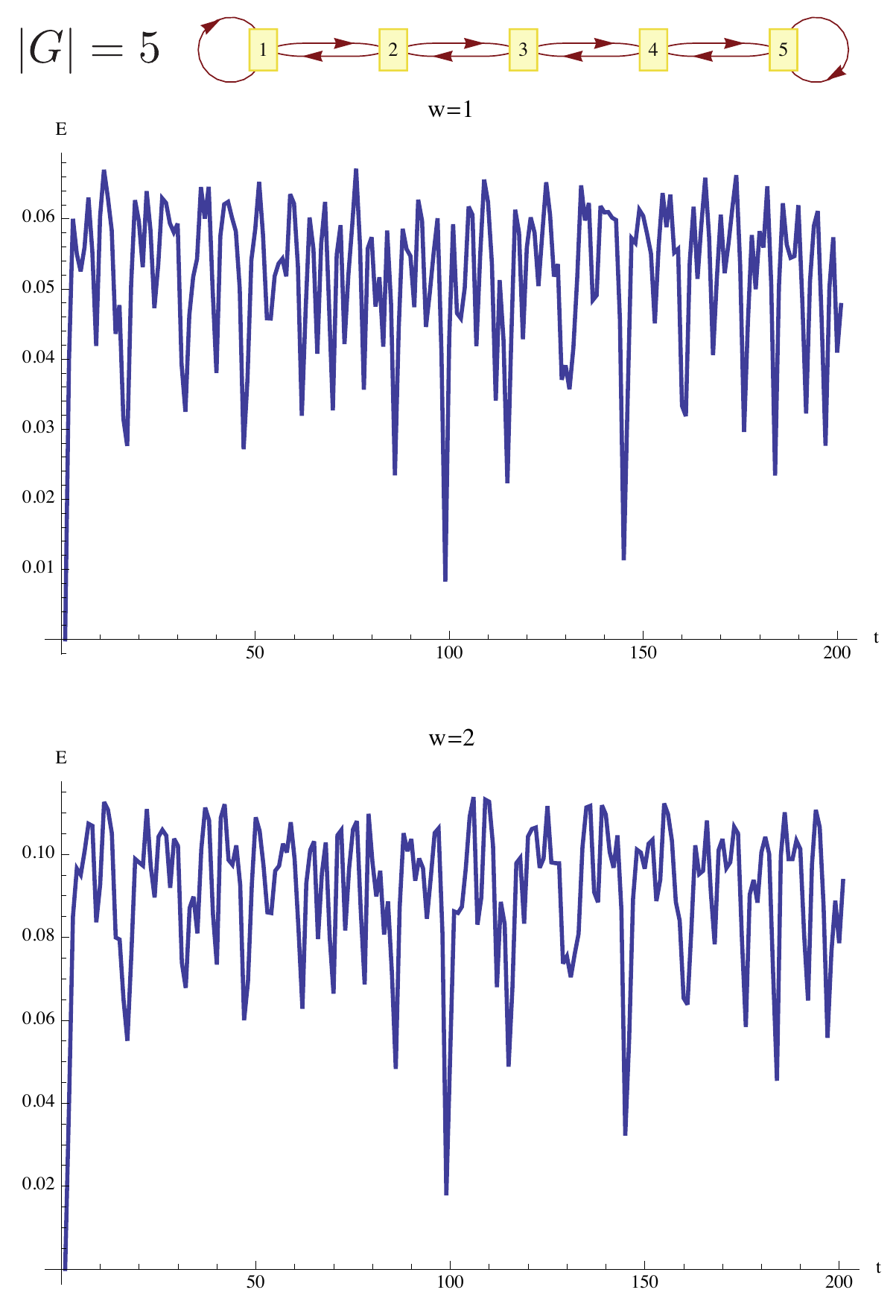}
\caption{(Colour online) The Meyer-Wallach entanglement measure for the one- (top) and two-walker (bottom) cases for a $|G|=5$ linear graph with $\delta=1/2$.} \label{fig:one_vs_two_walkers}
\end{figure}

Suppose we begin with two photons rather than one. Then the input state can be expressed as \mbox{$a_i^\dag a_j^\dag$}, where $i$ and $j$ denote distinct optical modes. Then, after some unitary transformation of the creation operators, the state of the system will be of the form \cite{bib:RohdeSchreiber10}
\begin{eqnarray}
a_i^\dag a_j^\dag &\to& \Big(\sum_k U_{ik} a_k^\dag\Big)\Big(\sum_l U_{jl} a_l^\dag\Big) \nonumber \\
&=& \sum_{kl} U_{ik} U_{jl} a_k^\dag a_l^\dag \nonumber \\
&=& \sum_{kl} M_{ijkl} a_k^\dag a_l^\dag.
\end{eqnarray}
Thus the transformation is a bilinear map from the initial two photon state to a superposition of two photon states. By linearity, the coefficients of any superposition of different two-photon input states will be transformed linearly. Applying the same ideas as before, taking the eigenvalues of $M$ will yield an eigenvalue spectrum where each of the eigenvalues gives a phase evolution of the corresponding eigenstate. Thus, quasi-periodicity may emerge. As an example comparison Fig. \ref{fig:one_vs_two_walkers} shows the Meyer-Wallach entanglement measure for the one- and two-walker cases with \mbox{$|G|=5$} and \mbox{$\delta=1/2$}. Evidently, while distinct, the two graphs share many of the main features. Applying the same techniques as before, one could calculate the eigenvalue spectrum of the unitary matrices to determine the overall frequency components in the dynamics. However, as the number of walkers increases the size of the unitary matrix grows exponentially as will the number of eigenvalues. Thus, the dynamics will in general become more complex as the number of walkers increases.

Next consider the situation where post-selection is performed. That is, we post-select on measuring a single photon at one output mode and consider the dynamics of the remaining photon. Upon measuring a single photon at mode $m$ the post-selected state of the system is given by \cite{bib:RohdeSchreiber10}
\begin{equation}
a_m^\dag \Big(U_{im} \sum_l U_{jl} a_l^\dag + U_{jm} \sum_k U_{ik} a_k^\dag\Big) = a_m^\dag \sum_k M'_{ijk} a_k^\dag.
\end{equation}
Thus, we are left with a product state where one photon is projected into mode $m$ and the remaining photon undergoes a linear map characterised by \mbox{$M_{ijk}' = U_{im}U_{jk} + U_{jm}U_{ik}$}.

\section{Conclusion \& future work}

In summary, we have considered the dynamics of an entanglement metric in discrete QWs. While we have focused on bounded linear QWs with a single walker, the walker operator formalism adopted from Ref. \cite{bib:RohdeSchreiber10} can easily be employed to generalise these results to arbitrary graph structures with any number of walkers. Our results are particularly relevant to present day experiments, where linear walks are considered, but obviously infinite graphs are not possible.

Our simple examples illustrate that highly complex, quasi-periodic dynamics can emerge even in small QW systems with boundary conditions. Interestingly, in the case of our simple walk on a line, it is evident that the boundary conditions are pivotal in the appearance of irregular dynamics, as similar studies on infinitely long lines demonstrate much more regular dynamics. Furthermore, even with the introduction of multiple walkers and post-selection, non-linear evolution cannot emerge.

We demonstrated that the walker operator formalism upon which our study is based, can easily be mapped to a linear optics photonic system. Thus, while challenging, experimental observation of irregular, quasi-periodic dynamics in linear optics systems is realistic with present-day technology. Indeed, with the addition of boundary conditions, recently demonstrated experimental QWs \cite{bib:Schreiber10,bib:Broome10,bib:Peruzzo10} could demonstrate such dynamics.

In some previous studies \cite{bib:VenegasAndraca09} it was observed that entanglement increases monotonically with time. However, our results indicate that in the bounded case this is not true and complex dynamics emerge. Thus, if the goal is to generate the maximal degree of entanglement, e.g. as a resource for other protocols, simply waiting for a very long time is not necessarily the best option. Therefore, timing is critical and understanding the precise nature of the entanglement dynamics is of importance.

Future work should give consideration to potential uses for the generalised W-type entanglement that is produced by QW systems. Considering more elaborate graph structures and alternate coin types would be valuable. This is something that can easily be approached using the walker operator formalism. In particular, the question of whether this can be used for efficient universal quantum computation or other quantum information processing applications is of importance.

\begin{acknowledgments}
We acknowledge support from the Australian Research Council. We give special thanks to Cathy Holmes, Gerard Milburn and Karen Dancer for helpful discussions.
\end{acknowledgments}

\bibliography{notes}

\end{document}